\documentclass[conference]{IEEEtran}

\IEEEoverridecommandlockouts
\usepackage{url}
\usepackage[noadjust]{cite}
\usepackage{amsmath,amsfonts,amssymb}
\usepackage{graphicx}
\usepackage{algorithm}
\usepackage[shortlabels]{enumitem}
\usepackage{flushend}

\def\BibTeX{{\rm B\kern-.05em{\sc i\kern-.025em b}\kern-.08em
		T\kern-.1667em\lower.7ex\hbox{E}\kern-.125emX}}

\begin{document}

\title{Disruption Prediction in Fusion Devices through Feature Extraction and Logistic Regression}

\author{
	\IEEEauthorblockN{Diogo R. Ferreira}
	\IEEEauthorblockA{\textit{IST, University of Lisbon, Portugal}\\
		diogo.ferreira@tecnico.ulisboa.pt}}

\maketitle

\thispagestyle{plain}
\pagestyle{plain}

\begin{abstract}
	This document describes an approach used in the \emph{Multi-Machine Disruption Prediction Challenge for Fusion Energy by ITU}, a data science competition which ran from September to November 2023, on the online platform Zindi. The competition involved data from three fusion devices -- C-Mod, HL-2A, and J-TEXT -- with most of the training data coming from the last two, and the test data coming from the first one. Each device has multiple diagnostics and signals, and it turns out that a critical issue in this competition was to identify which signals, and especially which features from those signals, were most relevant to achieve accurate predictions. The approach described here is based on extracting features from signals, and then applying logistic regression on top of those features. Each signal is treated as a separate predictor and, in the end, a combination of such predictors achieved the first place on the leaderboard.
\end{abstract}

\section{Introduction}

The goal of making nuclear fusion become a viable energy source is being pursued by public funding initiatives -- of which ITER~\cite{Claessens_2023} is currently the largest and most ambitious project -- and also by a series of private startup companies~\cite{Leslie_2022}, using a variety of different devices. One of the most common types of device is the \emph{tokamak}~\cite{Wesson_2011}, a toroidal machine where a plasma is heated up to millions of degrees, while being confined by strong magnetic fields. In tokamaks, there is the need to drive an eletric current through the plasma, and this current is often the source of instabilities~\cite{Blank_2008}. In an extreme case, the plasma might become so unstable that it \emph{disrupts}, i.e.~the particle confinement is lost, and the experiment is brought suddenly to an end. Of course, this also brings the plasma current to a halt; in a fraction of a second, the current can drop from a few mega amps to zero (the so-called \emph{current quench}~\cite{Riccardo_2004}), creating induced currents and extreme forces in the structure of the machine. For this reason, disruptions should be avoided and, in recent years, there has been a considerable effort in using machine learning techniques to predict disruptions before they occur~\cite{Rea_2018,Harbeck_2019,Vega_2022}.

The problem of disruption prediction in fusion devices is made more challenging by the fact that, in some cases, there is little or no training data available. In the case of ITER, for example, the requirements are quite stringent, as only 1\% of experiments will be allowed to end with a disruption~\cite{Strait_2019}; and yet, the machine is still under construction, so no operational data is available to train a disruption predictor. Therefore, the only option is to learn how to predict disruptions based on data from other machines. One of the earliest attempts at training a model on one machine and testing it on another reported that, while the model was able to achieve 90\% accuracy on the same machine it was trained on, it could not get past 70\% on a different machine~\cite{Windsor_2005}. Currently, the problem of transfer learning between disruption predictors across different machines is an active topic of research~\cite{Zheng_2023}.

In this context, the IAEA (International Atomic Energy Agency)\footnote{\url{https://www.iaea.org/}} together with the MIT Plasma Science and Fusion Center (PSFC),\footnote{\url{https://www.psfc.mit.edu/}} in collaboration with the Huazhong University of Science and Technology (HUST)\footnote{\url{https://english.hust.edu.cn/}} and the Southwestern Institute of Physics (SWIP)\footnote{\url{https://www.swip.ac.cn/}} in China, and under the auspices of the \emph{AI for Good} initiative of the International Telecommunication Union (ITU),\footnote{\url{https://aiforgood.itu.int/}} launched the \emph{Multi-Machine Disruption Prediction Challenge for Fusion Energy by ITU}, an online data science competition, which ran on the Zindi platform,\footnote{\url{https://zindi.africa/}} from late September to mid-November 2023.
		
The competition provided data from three machines (tokamaks): C-Mod, HL-2A, and J-TEXT. The goal was to use the training data from HL-2A and J-TEXT, together with a smaller set of training data from C-Mod, to develop a prediction model to be tested on a set of C-Mod experiments (referred to as \emph{shots}); the model was asked to predict whether each of those test shots is disruptive or not. The prediction accuracy was evaluated on the basis of the $F_1$-score, and submissions by participants were ranked on a public leaderboard. 

This report describes the approach that achieved the first place on the leaderboard. Basically, it extracts features from signals, and then applies logistic regression to learn a prediction model from each signal in the C-Mod training data. However, not all of these signals will have the same predictive power -- and this is where data from the other devices come into play. By applying the same approach on HL-2A and J-TEXT, for which larger datasets were available, it is possible to identify which signals are most promising for disruption prediction. Out of several possible choices, certain subsets of signals performed equally well or better than others, and those were the ones selected for scoring in the competition.

The following sections explain the rationale of the approach, and discuss its implementation and results.

\section{Feature Extraction}

\begin{figure}[h]
	\centering
	\includegraphics[width=\linewidth]{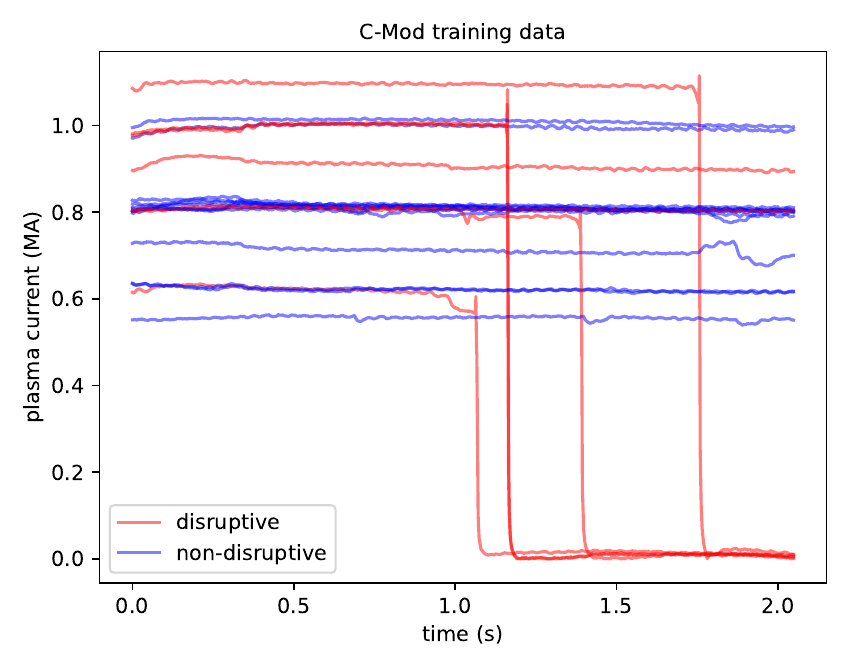}
	\caption{Plasma current for C-Mod training shots.}
	\label{fig_plasma_current}
\end{figure}

As mentioned in the introduction, one of the most obvious signs of a disruption is a sudden drop in the plasma current (the so-called \emph{current quench}). Although the current quench cannot serve as a good disruption predictor (because, by the time the current quench is observed, the disruption is already occurring), it does provide one of the most illustrative examples of a signal feature: in this case, the current quench is a feature (which can be defined, for example, as a steep gradient) of the plasma current signal. Figure~\ref{fig_plasma_current} illustrates that, for the C-Mod training data, the current quench is clearly visible in some disruptive shots.

It is possible to imagine that an experienced researcher in this field will know about many signals and features which are related to disruptions, and which could be used for disruption prediction. For example, it is known that core radiation (i.e.~excessive radiation from the core region of the plasma) is often the cause of disruptions~\cite{Rossi_2023}, and this can be measured by X-ray diagnostics. As another example, it is well-known that certain magneto-hydro-dynamic (MHD) instabilities, such as the locked mode~\cite{Sias_2019}, often lead to disruptions, and these can be detected based on magnetics. So an experienced researcher might know exactly where and what to look for, in search for signs of an impending disruption.

On the other hand, the vast set of diagnostics and signals available from each machine creates the opportunity to identify these and other patterns that could be relevant to disruption prediction. In particular, which features from which signals would be helpful to find out whether a disruption is about to occur? A data scientist should be able to answer this question by training a machine learning model on a dataset that includes both the signals from each experiment and a label indicating whether that experiment ended in a disruption or not. This is precisely what the competition provided, for three different machines, with similar diagnostic systems.

Now, without knowing the patterns of disruptive behavior beforehand, a data scientist could try to generate as many features as possible from each signal, and then train a model to learn how those features correlate with the target variable (which, in this case, is a binary label indicating whether the shot is disruptive or not). For the purpose of generating features from signals, we used the \texttt{tsfresh} library,\footnote{\url{https://tsfresh.readthedocs.io/}} a Python library that extracts features from time series. Among the wide range of features that \texttt{tsfresh} can extract, we find:
\begin{itemize}

	\item basic statistics, such as minimum, maximum, mean, median, etc.;
	
	\item aggregate statistics, such sum of values, absolute sum of consecutive changes, etc.
	
	\item descriptive statistics, such as those based on autocorrelation, linear regression, etc.
	
	\item counting statistics, such as the number of values above or below the mean, number of peaks, crossings, etc.
	
	\item positional statistics, such as the location of maxima and minima across the time series;
	
	\item spectral statistics, such as the fast Fourier transform (FFT) coefficients, mean and variance of the spectrum, etc.
	
\end{itemize}

This large assortment of features is probably too much to consider, and some of them may be overly complicated for the problem at hand (in the sense that they may go beyond what a physicist would be able to recognize by looking at the plot of a signal). For this reason, we restrict the generated features to those that can be computed efficiently, and avoid some computationally expensive features that would bring little added value. Still, we ended up with more than 700 features. This is the amount of features that are extracted from each signal, regardless of the length and sampling rate of the signal (which is important, because the length and sampling rate may vary across signals, experiments, or machines).

After computing the features, but before training a model, we normalize them across the dataset, by subtracting the mean and dividing by the standard deviation. This step is recommended, as different features might have different scales. Another step is to remove the features with very low or negligible variance; this is often due to features with a constant value across the training dataset.

\section{Logistic Regression}

As described in the previous section, we extract a large number of features from each signal (e.g.~the plasma current, or any other signal that is available in a shot). It is those features that we use to train machine learning models. In fact, we train a separate model for each signal, i.e.~each signal is used as a separate disruption predictor. By collecting the same signal across multiple shots, together with the target label for each shot, we can train a machine learning model to predict the target label for a new shot which has that signal.

In the case of plasma current, it is expected that every shot should have this signal. However, that is not the case with other signals; sometimes, a signal exists in some shots but not in others. Therefore, each predictor is trained on the set of shots where the signal is available. Conversely, a predictor can only applied on shots where the signal is available.

Naturally, a prediction based on a single signal might not be very accurate; however, since each shot has multiple signals, the idea is that a more accurate prediction can be achieved by combining multiple predictors, where each predictor corresponds to a different signal. Since the final prediction is the result of combining multiple predictors (an ensemble), and since the training set for each predictor is limited to the shots that contain the corresponding signal (limited dataset size), there is no strong motivation to use very sophisticated models, or models that could severely overfit the training data.

In contrast, a simple type of model -- namely logistic regression~\cite{Murphy_2022} -- should suffice to provide an output between 0 and 1 for a given set of input features. If $\boldsymbol{x}\!=\![x_1, x_2, ..., x_k]$ is a vector of $k$ features extracted from a signal, and $p$ is the estimated probability of disruption, then:
\begin{equation}
	p = \frac{1}{1+e^{-(\boldsymbol{w}\cdot\boldsymbol{x}+b)}}
	\label{eq_prob}
\end{equation}
where $\boldsymbol{w}\!=\![w_1, w_2, ..., w_k]$ (a vector of weights) and $b$ (a bias) are parameters to be learned during training by minimizing the binary cross-entropy over a set of $n$ training samples:
\begin{equation}
	\mathrm{loss} = -\frac{1}{n} \sum_{i=1}^{n} \left[ y_i \log p_i + (1-y_i) \log (1-p_i)\right]
	\label{eq_loss}
\end{equation}
where $y_{i}$ is the true label of sample $i$, and $p_i$ is the estimated probability of disruption for that sample.

This is precisely the logistic regression model provided by \texttt{scikit-learn},\footnote{\url{https://scikit-learn.org/}} provided that one is careful to set the penalty (regularization) parameter to \texttt{None}, to avoid adding a regularization term to the loss function in Eq.~(\ref{eq_loss}).

\section{Implementation}

Now that both stages -- feature extraction and logistic regression -- have been explained, Algorithm~\ref{alg_predictions} provides an overview of the implementation, where each signal is the basis for training a classifier (on the training shots that contain the signal) and for making predictions (on the test shots that contain the signal). If a training shot does not contain the signal, then that shot will not be used to train the classifier (but it will be used to train other classifiers, according to the signals that it contains). Likewise, if a test shot does not contain the signal, then the classifier will not be used to make a prediction on that shot (but other classifiers will, according to the signals that the shot contains). In the end, each test shot will have multiple predictions, from multiple classifiers, according to the signals that it contains.

\begin{algorithm}
	\caption{Overview of model training and predictions}
	\label{alg_predictions}
	\vspace{0.5em}
	For each machine, do the following:
	\begin{enumerate}[1.]
		\item Read the data from all shots on that machine.
		\item Split the shots into a training set and a test set, according to the following options:
		\begin{enumerate}[a)]
			\item if the machine is C-Mod, the training shots and the test shots are as specified in the competition data;
			\item if the machine is HL-2A or J-TEXT, split the shots randomly into 50\% for training and 50\% for testing.
		\end{enumerate}
		\item Find the set of all signals available in the training shots (e.g.~plasma current, magnetic field, etc.).
		\item For each signal, do the following:
		\begin{enumerate}[a)]
			\item collect all the shots that have the signal, from both the training set and the test set;
			\item extract the signal features from each of those shots;
			\item normalize each feature across all of those shots;
			\item remove any features with very low variance;
			\item train a logistic regression classifier on the training shots that have the signal;
			\item use the classifier to predict the probability of disruption on the test shots that have the signal;
		\end{enumerate}
		\item Save the predictions of each classifier for each test shot.
	\end{enumerate}
	\vspace{0.5em}
\end{algorithm}

For the competition, a single prediction is required for each test shot. One of the simplest ways to generate such submission is to average the multiple predictions available for each shot. This already provides a good result, but it is possible to do even better by selecting a subset of signals as predictors. To investigate this, we carried out experiments with HL-2A and J-TEXT data, where (as indicated in Algorithm~\ref{alg_predictions}, step 2) we used 50\% of shots for training and 50\% for testing. The results from those experiments suggest that the predictive accuracy of each classifier, when taken individually, and as measured by the $F_1$-score, can vary in the range 0.85--0.95 for J-TEXT, and 0.80--0.97 for HL-2A.

For C-Mod, the prediction accuracy for any individual classifier is expected to be lower, because the training set is much smaller (only 20 training shots for C-Mod, compared to about 500 for HL-2A, and about 1000 for J-TEXT). Therefore, it would be important to focus on the best-performing predictors to improve the overall accuracy, while avoiding the weakest classifiers which could bring spurious predictions (i.e.~noise) to the results, worsening the accuracy.

To identify which signals/predictors are the most promising, we again turned to J-TEXT and HL-2A. For those machines, if we rank the classifiers according to their individual accuracy, we find that some magnetics signals are at the top of the list, followed by a few radiation measurements, and other signals interspersed in between. In some cases, it was possible to identify the corresponding signals in C-Mod; in other cases, especially in multi-channel diagnostics such as radiation and magnetics, it was harder to find the exact correspondence, so we tried different choices, while observing that they often produced the same results in terms of scoring.

So, on one hand, the idea is to reduce the set of signals to the most promising ones, and this can be done by looking at the most promising signals on the other machines. On the other hand, we should try to keep several predictors for each test shot, because, in light of the small number of training shots available for C-Mod, no single predictor would be expected to be as good as in the case of the other machines.

An $F_1$-score in the range of what was obtained for J-TEXT and HL-2A, where a single predictor could reach 0.95 or more, is probably out of reach for C-Mod. In any case, we tried to find a balance between having too many signals (which might include weak predictors) and having too few signals (which might not include enough predictors). In addition, the multi-channel diagnostics  (e.g.~radiation, magnetics) should not be over-represented; from these diagnostics, only a few channels should be included, so as not to dilute the predictive power of other, possibly stronger predictors.

In the end, we selected the following set of signals (where some signal names have been identified from~\cite{Mathews_2019} and others from the documentation provided for the competition):
\begin{itemize}
	\item the line integral density (center chord);
	\item the Greenwald fraction;
	\item three radiation emission channels (e.g.~3, 9, 23);
	\item three magnetics signals (e.g.~2, 13, 20);
	\item the toroidal magnetic field;
	\item the vertical elongation ($\kappa$);
	\item the internal inductance ($l_i$);
	\item the edge safety factor ($q_{95}$);
	\item the horizontal and vertical displacements;
	\item the loop voltage ($V_{\mathrm{loop}}$);
	\item the stored plasma energy ($W_{\mathrm{plasma}}$);
	\item five soft X-ray radiation channels (e.g.~2, 14, 33, 36, 38);
	\item the locked mode (LM) proxy.
\end{itemize}

Overall, we selected about 20 signals from a universe of about 170 signals. Yet, this should be regarded as an example rather than a definitive list, since a different selection of signals may yield the same or even better score. This particular selection, when applied to the results of Algorithm~\ref{alg_predictions}, achieved an $F_1$-score of approximately 0.937 on the public leaderboard, and 0.964 on the private leaderboard.\footnote{Based on the true labels that were provided at the end of the competition, this corresponds to a TP rate of 0.964, TN rate of 0.974, FP rate of 0.025, FN rate of 0.036, precision of 0.950, recall of 0.964, $F_1$-score of 0.956, and area under the ROC curve (AUC) of 0.980, on the test set (public + private).}

\section{Discussion}

At the end of this competition, it would be interesting to know what our predictors have learned from those signals, and which features turn out to be important for disruption prediction. By looking at the coefficients of each trained classifier, i.e.~the learned weights $\boldsymbol{w}$ in Eq.~(\ref{eq_prob}), it is possible to identify which features have the largest (absolute) weight. Since features have been normalized, such weights can provide a measure of the relative importance of each feature.

When analyzing the most important features for each classifier, there are some curious findings:
\begin{itemize}

	\item For some signals, the most important feature is the data length. Naturally, there is a certain tendency for disruptive shots to be shorter, so data length is inversely related to disruptivity. This feature often appears associated with signals that come from multi-channel diagnostics (i.e.~radiation, magnetics). In a way, this might explain why the channels from those diagnostics seemed to be somewhat interchangeable, if data length is what matters the most.
	
	\item Some signals have very specific features. For example, one of the magnetics signals has, as its most important feature, the energy ratio of the last of 10 segments (this means cutting the signal into 10 segments, and computing the sum of squares of the last segment divided by the sum of squares of the entire signal). There is probably some disruptive MHD behavior being detected in this signal.
	
	\item For the toroidal magnetic field, the most important feature is related to the location of its maximum. It seems that, for disruptive shots, there are small blips or peaks in the magnetic field later in the shot.
	
	\item For the stored plasma energy, the most important feature is the number of peaks of support 50 (i.e. the number of peaks that are separated from other peaks by at least 50 data points; at the sampling rate of 1 kHz, this corresponds to 50 ms).

	\item For the locked mode amplitude, the most important feature is the longest strike above the mean (i.e. the length of the longest stretch above the mean of the signal). This is not surprising if we consider that, when a locked mode appears, there is a jump in the locked mode signal, and a disruption is almost certain to occur shortly afterwards.
	
	\item For some signals, the most important features are related to specific coefficients of the fast Fourier transform (FFT). This means that the presence or absence of certain frequencies in the spectrum might be related to disruptivity, as is the case with the loop voltage, for example.
	
	\item For other signals, namely vertical elongation, the most important feature is the correlation of the variance of a certain segment to a linear regression over the variances of previous segments.
	
	\item For several signals (including internal inductance, edge safety factor, and horizontal displacement) the most important features are related to specific coefficients of an autoregressive fit. This autoregressive fit measures the extent to which the current value of a signal can be predicted based on its past values. When such coefficient is lower, the signal is more unpredictable, and the shot is more likely to be disruptive.

	\item Finally, there is the curious case of the soft X-ray radiation channels, where one channel uses the correlation to a linear regression over previous segments, another channel uses the data length, a third channel uses an FFT coefficient, and the last two channels use coefficients of an autoregressive fit.

\end{itemize}

This brief description only scratches the surface, since there are many more features that the classifiers make use of, with slightly less but still comparable weight. However, our goal here was just to provide a general idea of the wide range of features that might turn out to be useful to identify patterns of disruptive behavior in these diagnostic signals. 

\bibliographystyle{IEEEtran}
\bibliography{report}

\end{document}